# A Response Embedded Atom Method of Interatomic Potentials

L.G. Zhou and Hanchen Huang*

Department of Mechanical Engineering, University of Connecticut, Storrs, CT 06269

The embedded atom method (EAM) potentials are probably the most widely used interatomic potentials for metals and alloys. However, the EAM potentials impose three constraints on elastic constants that are inconsistent with experiments. At a more subtle (but more important) level, the EAM potentials often incorrectly describe the outward/inward relaxation of surface layers, and therefore will not reliably describe nanostructures. This Letter reports a response EAM (R-EAM) that addresses both issues. Conceptually, the electron distribution from each atom does not respond to the atom's environment within the EAM. In reality, the electron distribution from each atom depends on the atom's environment, and this dependence is explicitly incorporated in the R-EAM. Analytical derivation shows that the R-EAM potentials do not impose these three constraints on elastic constants that EAM potentials do. Further, taking hexagonal close packed (HCP) metals Ti, Mg, and Zn as the prototypes, the authors show that the R-EAM potentials correctly describe surfaces – in terms of interlayer spacing and surface reconstruction, in agreement with quantum mechanics calculations, while EAM potentials are not in agreement. In comparison to EAM potentials, the R-EAM potentials require only approximately twice the computational power.

* Author to whom correspondence should be made; electronic mail: hanchen@uconn.edu

Interatomic potentials are the foundation of classical molecular mechanics simulations (including statics and dynamics simulations), and they dictate whether the simulation results are physically reliable. In comparison to quantum mechanics calculations, the classical molecular mechanics simulations are capable of capturing longer time evolution of larger physical systems. For example, the dynamics of dislocation interactions and surface diffusions can be too challenging computationally for quantum mechanics calculations, but they are well within the reach of classical molecular dynamics simulations. With the widespread use of molecular mechanics simulations, multiple types of interatomic potentials have been developed, and they include (a) pair-wise potentials [1,2] for condensed matters of closed-shell atoms, (b) embedded-atom-method (EAM) potentials [3] and similar formulations [4-6] for metals and alloys, (c) bond order potentials [7] for covalent solids, and (d) force fields [8] for organic and oxide systems.

For metals and alloys, the EAM potential is probably the most widely used approach in the research literature. Three other formulations are effectively similar to the EAM potential and they are the effective medium theory [4], Finnis-Sinclair potential [5], and the glue model [6]. Classical molecular mechanics simulations that are based on EAM potentials are prolific; for example, a recent review [9] offers a glimpse of the wide range applications. Due to the more frequent use of EAM potentials than the other three approaches, in the following we take the EAM to first identify issues, then propose a solution – a response EAM (R-EAM), and finally demonstrate advantages of the R-EAM over the EAM.

EAM potentials have two major advantages, relative to pair potentials, and they are: (a) the unphysical constraint on elastic constants $C_{12}=C_{44}$ is eliminated, and (b) bond saturation and therefore surface bonding is better represented. Along the same lines, EAM potentials also have two remaining issues to be addressed. As the first issue, EAM potentials still impose three other constraints on elastic constants, and they are (a) $C_{12}>C_{44}$ for cubic crystals [9, 10], (b) $C_{13}>C_{44}$ and (c) approximately $3C_{12}-C_{11}>2(C_{13}-C_{44})$ for hexagonal close packed (HCP) crystals [10, 11]. However, experimental data are inconsistent with these three constraints. For example, experimental values of elastic constants of Cr are $C_{12}=68GPa$ and $C_{44}=101GPa$ [12]; i.e., $C_{12}<C_{44}$. As the second example, the experimental values of elastic constants of Be are $C_{13}=6GPa$ and $C_{44}=163GPa$ [12]; i.e., $C_{13}<C_{44}$. Finally as the third example, the experimental values of elastic constants of Zn are $3C_{12}-C_{11}=-65GPa$ and $2(C_{13}-C_{44})= 20GPa$ [13]; i.e., $3C_{12}-C_{11}<2(C_{13}-C_{44})$.

As the second issue, EAM potentials usually do not correctly describe the outward/inward relaxations of surface layers [14], and therefore will not reliably describe nanostructures. For example, EAM-based simulations contradict experimental observations on the relaxation of A1{111} [15].

In this Letter, we propose the R-EAM that eliminates the three constraints on elastic constants and captures the outward/inward relaxation of surface layers, in contrast to the inability of EAM in these cases. To introduce the concept of a response function, we start with the EAM formulation [3]:

$$E = \sum_i E_i = \sum_i F(\bar{\rho}_i) + \frac{1}{2}\sum_{ij(i\neq j)} \phi(r_{ij}) \text{ and } \bar{\rho}_i = \sum_{j(\neq i)} \rho(r_{ij}) \qquad (1)$$

Here, $E$ is the total energy of a system of atoms, $E_i$ is the nominal potential energy of atom $i$, $F$ is the embedding function, $\bar{\rho}_i$ is the total electron density at atom $i$ that includes electron contributions from other atoms, and $\phi(r_{ij})$ is the pair interaction energy between atoms $i$ and $j$. According to the EAM, the contribution of electron density at atom $i$ by atom $j$, $\rho(r_{ij})$, is independent of the environment of atom $j$. That is, whether atom $j$ is surrounded by two other atoms or five other atoms, the contribution $\rho(r_{ij})$ is the same. Physically, $\rho(r_{ij})$ should depend on the environment of atom $j$. To account for this dependence, we introduce a response function $\Re$ so that the contribution of atom $j$ to the electron density at atom $i$ becomes:

$$\rho_R(r_{ij}) = \rho(r_{ij})[1+\Re_j(\bar{\rho}_{ji})] \qquad (2)$$

The response function of atom $j$, $\Re_j$, is a function of its local density of electrons $\bar{\rho}_{ji}$ that are contributed from other atoms; to achieve a de-coupled expression, this density excludes contribution from atom $i$ or $\bar{\rho}_{ji} = \bar{\rho}_j - \rho(r_{ij})$. Directly using the expression of equation (2) in the potential will result in substantial increase in computational time. Within the framework of EAM, the partition of embedding energy and pair interaction energy is not unique [16]. As a result, it is possible that the same effects can come from another response function $R$ through the following expression:

$$E = \sum_i E_i = \sum_i F(\bar{\rho}_i) + \frac{1}{2}\sum_{i\neq j} \phi(r_{ij})\left[1+R(\bar{\rho}_{ij},\bar{\rho}_{ji})\right] \text{ and } \bar{\rho}_i = \sum_{j(\neq i)} \rho(r_{ij}) \qquad (3)$$

In passing, we note that the expression of R-EAM potentials in equation (3) is consistent with an approximate form of the density functional theory [17]. Incorporation of the response function in the form of equation (3) leads to substantial higher computational efficiency than in the form of equation (2). This is the form of the R-EAM potential that we propose and use in the following. Considering the symmetry of atoms $i$ and $j$ in the pair interaction, we further take $R(\bar{\rho}_{ij}, \bar{\rho}_{ji}) = R(\sqrt{\bar{\rho}_{ij}\bar{\rho}_{ji}})$. It is worth mentioning that the computational cost of molecular mechanics simulations based on R-EAM potentials in the form of equation (3) is only about twice of that based on the EAM potentials.

Having described the concept and the mathematical form of the R-EAM potential, we next demonstrate its numerical determination. For this purpose, we need to make three choices. The first choice pertains to materials. Since two of the three constraints on elastic constants are for HCP metals, we choose Ti, Mg, and Zn as the prototypes of different c/a ratios. Ti has a c/a ratio that is smaller than the ideal value of 1.633 (and it is more commonly used than Be that also has a smaller c/a ratio), Mg has a c/a ratio that is close to the ideal value, and Zn has a c/a ratio that is larger than the ideal value. Correct representation of relationships among elastic constants of HCP metals is the first test of the R-EAM; the EAM leads to contradiction with experimental results for materials such as Zn.

The second choice pertains to the balance of determination and fitting. For clarity, we will determine $\rho(r)$ as the electron density of an isolated atom, without fitting. We also determine the pair interaction energy $\phi(r)$ as a function of dimer distance while not allowing the electron density of isolated atoms to relax, without fitting. Both of these two determinations will use quantum mechanics calculation results as the reference. For transferability, we will fit the response function $R(\rho)$ and the embedding function $F(\rho)$ to a range of data for small clusters, interfaces, and bulk solids. Specifically, the quantum mechanics calculation results for the fitting of the two functions $R(\rho)$ and $F(\rho)$ in equation (3) are:

(a) The force on an atom in equilateral dimer, trimer, and tetramer. In each case, the equal distance of two atoms is variable between 0.20 and 0.65 nm, with the following 13 meshing points: 0.20, 0.22, 0.24, 0.26, 0.28, 0.29, 0.31, 0.33, 0.37, 0.42, 0.48, 0.55, and 0.65 nm;

(b) Stress tensors for face-centered-cubic, body-centered-cubic, simple cubic, and HCP crystal structures with various nearest neighbor distances. For each structure, the nearest neighbor distance is meshed in the same way as in (a);

(c) Stress tensors for two-dimensional equilateral triangular, square, and graphitic crystal structures with various nearest neighbor distances. For each structure, the nearest neighbor distance is meshed in the same way as in (a); and

(d) Formation energies of unrelaxed defect configurations including surfaces $\{0001\}$, $\{11\bar{2}0\}$, $\{10\bar{1}0\}$, $\{10\bar{1}1\}$; $\{10\bar{1}1\}$ twin boundary; $\{0001\}$ stacking fault; and vacancy.

In addition, we use experimental data of cohesive energy in HCP structure to calibrate the configuration energy from quantum mechanics results. Experimental data of lattice constants (c and a) and elastic constants $C_{ij}$ (at temperature T=0K or 4K) are also used in the fitting. Even though values of lattice constants are at room temperature, their differences with their zero temperature counterparts are less than 1% and within the fitting error.

The third choice pertains to the way of generating quantum mechanics results for potential determination and fitting. In order to rigorously represent electron density, we choose the accurate all-electron full-potential linearized augmented plane-wave method (FP-LAPW) [18] to generate the results in list (a) to (d); based on the numerical code ELK [19]. The Perdew-Burke-Ernzerhof generalized gradient approximation exchange-correction functional [20] is used. The cut-off length for plane waves in the interstitial region is $8.0/R_{min}$ for energy convergence of 0.01 eV. Here, $R_{min}$ is the smallest muffin-tin radius; and it is 0.10 nm for Mg and Zn, and 0.08 nm for Ti, to ensure full relaxation of a dimer. The number of k-points is chosen for each case to ensure that further increase of k points has no impact on the total energy, within the accuracy of 0.01 eV.

Following these three choices, we have determined the function $\rho(r)$ and $\phi(r)$, as shown in Figures 1(a) and 1(b). For each function, there are 100 data points, within the range of [0.10 nm, 0.65 nm]; between two neighboring data points, each function is in the form of a cubic spline. The cutoff function for both $\rho(r)$ and $\phi(r)$ is: $f(r) = 0.5 + 0.5\cos[20\pi(r-0.6)]$ with $r$ in nm; it becomes one as $r$ is smaller than 0.60 nm and zero as $r$ is larger than 0.65 nm.

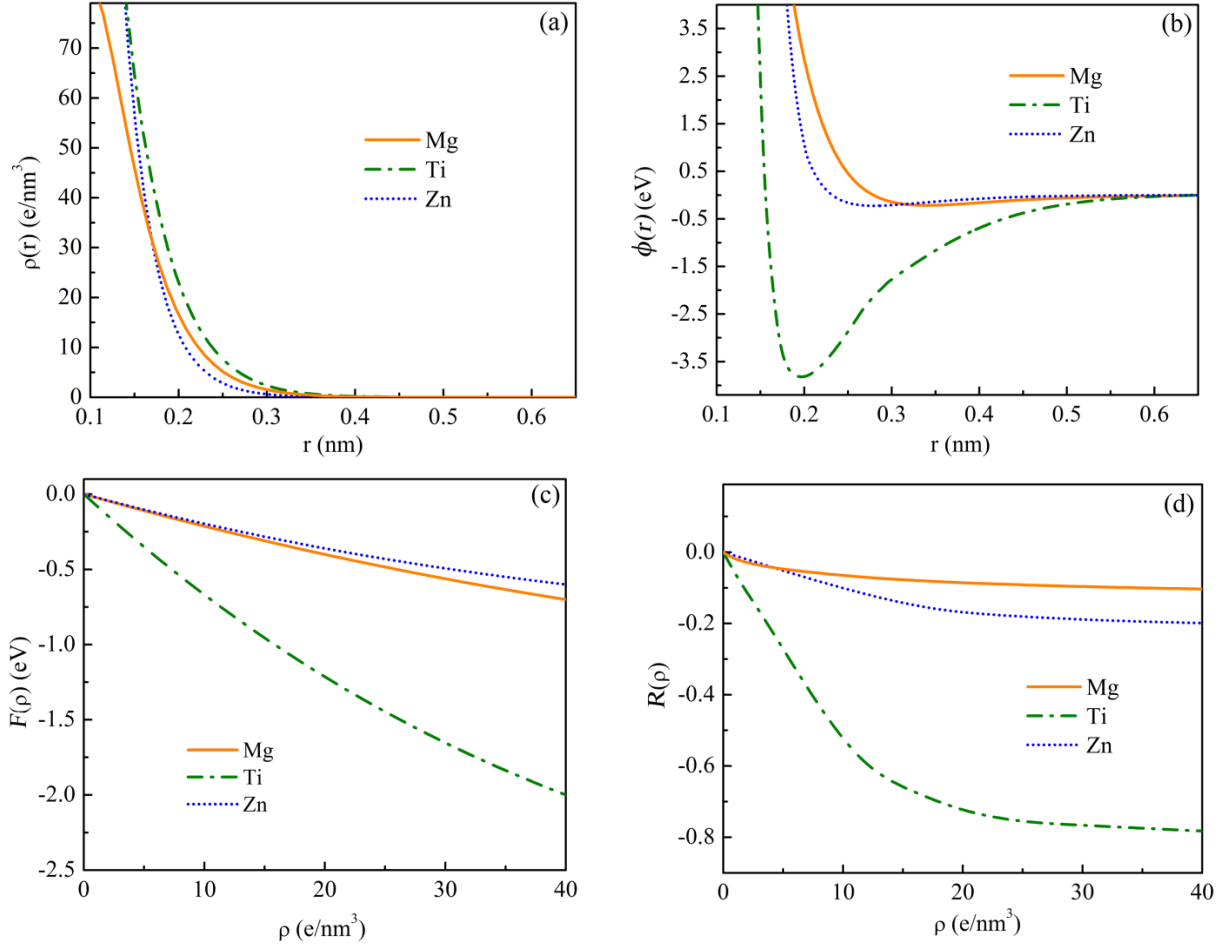

FIG. 1. Determined functions (a) $\rho(r)$, and (b) $\phi(r)$; and fitted functions (c) $F(\rho)$ and (d) $R(\rho)$, for the three HCP metals.

Table 1: Values of $F(\rho)$ and $R(\rho)$ at nine knot points for Mg, Ti and Zn

| | Mg | | Ti | | Zn | |
|---|---|---|---|---|---|---|
| $\rho$ (eV/nm$^3$) | $F$ (eV) | $R$ | $F$ (eV) | $R$ | $F$ (eV) | $R$ |
| 0 | 0.0000 | 0.0000 | 0.0000 | 0.0000 | 0.0000 | 0.0000 |
| 1 | -0.0212 | -0.0211 | -0.0734 | -0.0594 | -0.0215 | -0.0105 |
| 3 | -0.0648 | -0.0375 | -0.2162 | -0.165 | -0.0634 | -0.0324 |
| 6 | -0.1373 | -0.0524 | -0.4202 | -0.3256 | -0.1236 | -0.0634 |
| 12 | -0.2555 | -0.0717 | -0.7936 | -0.5940 | -0.2342 | -0.1250 |
| 16 | -0.3313 | -0.0791 | -1.0180 | -0.6732 | -0.3011 | -0.1571 |
| 20 | -0.4035 | -0.0863 | -1.2241 | -0.7227 | -0.3629 | -0.1729 |
| 30 | -0.5664 | -0.0966 | -1.6651 | -0.7662 | -0.4967 | -0.1964 |
| 40 | -0.7177 | -0.1042 | -2.0123 | -0.7821 | -0.6038 | -0.2077 |

For each of the two fitted functions $F(\rho)$ and $R(\rho)$, we use cubic splines with nine spline knot points. Table 1 lists the values of $F(\rho)$ and $R(\rho)$ at the nine knot points, and Figures 1(c) and 1(d) show their dependence on $\rho$. Table 2 compares the lattice constants, cohesive energy, and elastic constants based on the R-EAM potentials with their experimental counterparts. Having put the concept of R-EAM potential in numerical forms for three HCP metals, we now demonstrate the advantages of this method, in terms of the two remaining issues of EAM potentials that are outlined earlier.

As we recall, the first issue is that EAM potentials impose three constraints on elastic constants: (a) $C_{12}>C_{44}$ for cubic crystals, (b) $C_{13}>C_{44}$ and (c) approximately $3C_{12}-C_{11}>2(C_{13}-C_{44})$ for HCP crystals. We have analytically derived the elastic constants $C_{ij}$ based on R-EAM potentials and compared them with their counterparts based on EAM potentials [10]. The derivation and comparison: (1) clarify that the first constraint for cubic crystals always applies based on EAM, that the second and the third constraints for HCP crystals apply only when the electron density in the potential is a monotonic function; and (2) show that R-EAM potentials do not impose any of the three constraints – the numerical results in Table 2 (even for Zn) also confirm the derivation. That is, the proposed R-EAM does satisfactorily address the remaining issue of EAM in describing elastic constants.

Table 2: Comparison R-EAM results with experimental values (Exp) of lattice constants (a and c) and cohesive energies ($E_c$) from [21], elastic constants (B and $C_{ij}$) from [13].

|  | Mg | | Ti | | Zn | |
|---|---|---|---|---|---|---|
|  | R-EAM | Exp | R-EAM | Exp | R-EAM | Exp |
| a (nm) | 0.321 | 0.321 | 0.295 | 0.295 | 0.265 | 0.266 |
| c/a | 1.623 | 1.623 | 1.587 | 1.586 | 1.854 | 1.861 |
| $E_c$ (eV) | 1.51 | 1.51 | 4.85 | 4.85 | 1.35 | 1.35 |
| B (GPa) | 37 | 37 | 110 | 110 | 80 | 80 |
| $C_{11}$ (GPa) | 64 | 63 | 176 | 176 | 176 | 179 |
| $C_{12}$ (GPa) | 25 | 26 | 82 | 87 | 44 | 38 |
| $C_{13}$ (GPa) | 22 | 22 | 72 | 68 | 55 | 55 |
| $C_{33}$ (GPa) | 67 | 66 | 187 | 191 | 63 | 69 |
| $C_{44}$ (GPa) | 18 | 18 | 50 | 51 | 49 | 46 |

As the second advantage of R-EAM over EAM method in describing surfaces, we are not able to show it analytically; rather, we will be able to show this advantage numerically. For this purpose, we need a reference from best performing EAM potentials and another reference from quantum mechanics calculations for each case. In choosing EAM potentials, our criterion is that they well or best describe surfaces and interfaces; they are the potential by Zope and Mishin for Ti [16] and that by Liu et al. for Mg [22], respectively. For Zn, only two available EAM potentials stabilizes the HCP structure at stress free condition (instead of pressure free condition), and we choose the one that better describes elastic constants [23]. Calculations based on this EAM potential [23] are not the fair reference for comparison with those based on our R-EAM potential, but they are included anyway for completeness.

The second reference necessary for the comparison is a database of quantum mechanics calculations. Here, complete details of electron density are not necessary but simulations of sufficiently large systems are. As a tradeoff, the local orbital based pseudo-potential density functional code SIESTA [24] is chosen to establish the reference. The exchange-correlation functional is given by the Ceperley-Alder local density approximation [25]. Further, for contrast, we have included calculation results based on the best performing EAM potentials available for each HCP metal according to its description of elastic constants [16,22,23]. The quantum mechanics calculations cover five high symmetry surfaces $\{0001\}$, $\{10\bar{1}0\}$, $\{11\bar{2}0\}$, $\{10\bar{1}1\}$, $\{10\bar{1}2\}$. Other than $\{0001\}$ and $\{11\bar{2}0\}$ surfaces, the rest are in the form of double layers that are not separated. As a convention, the interlayer spacing involving double layers refers to the distance between the bottom of upper double-layer and the top of the lower double-layer.

Comparing R-EAM and EAM based calculations with quantum mechanics calculations, we now demonstrate the second advantage of the R-EAM method. Figure 2a shows that the R-EAM potential reproduces surface outward relaxation or increase of the first interlayer spacing, in agreement with the quantum mechanics results; in contrast, the EAM potential gives rise to inward relaxation. Figure 2b further shows that the agreement between R-EAM based and quantum mechanics calculations is universal to all the five surfaces for each of the three HCP metals, in contrast to the disagreement between EAM based and quantum mechanics calculations.

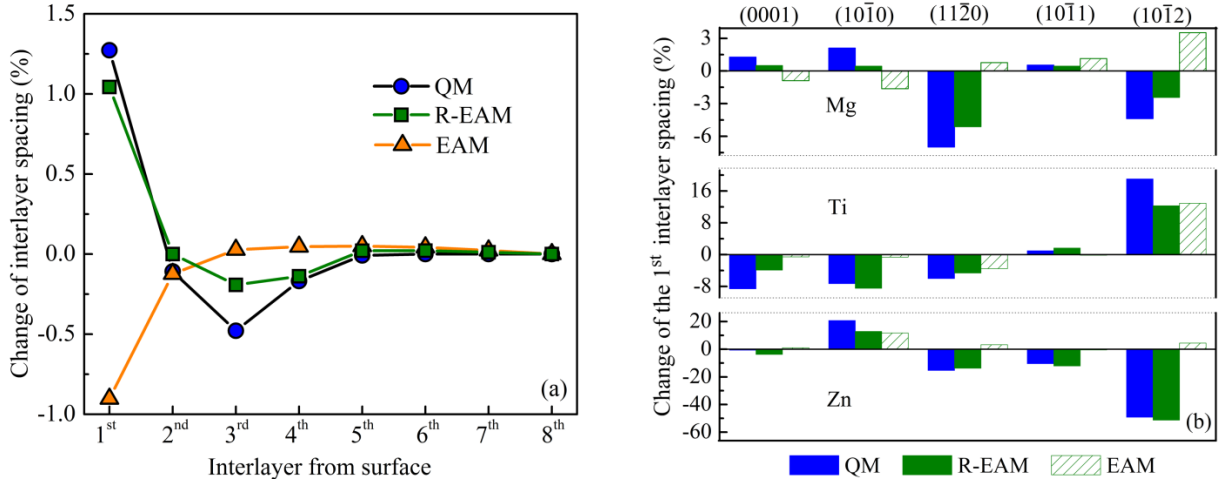

FIG. 2. (a) Change of interlayer spacing – relative to bulk values – for Mg {0001}, according to quantum mechanics (QM), R-EAM, and EAM based calculations; (b) Change of the first interlayer space for Mg, Ti and Zn, according to QM, R-EAM, and EAM based calculations.

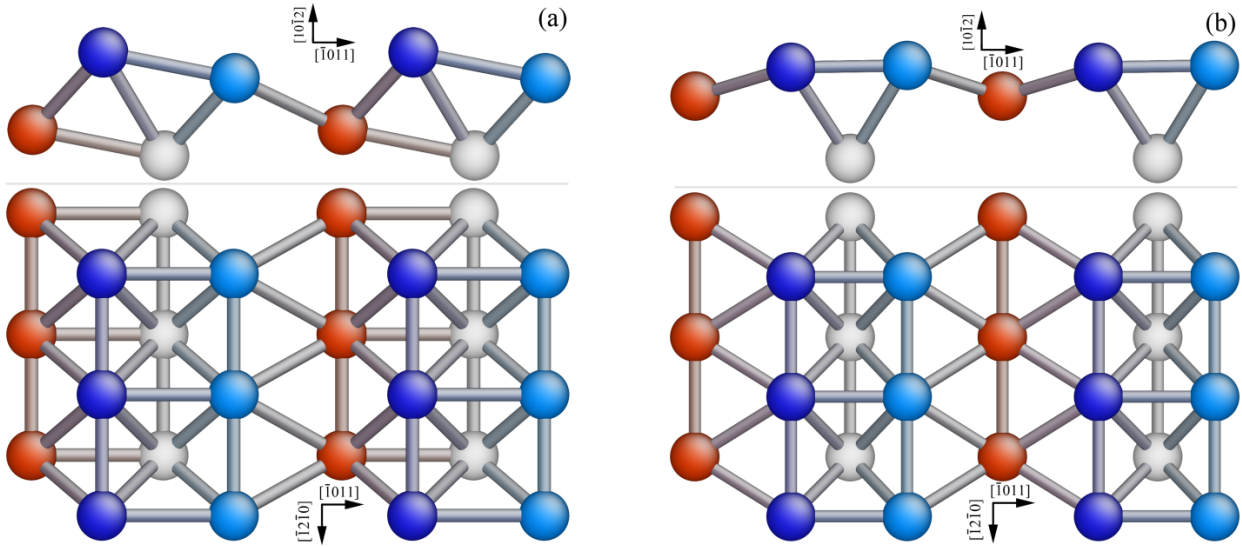

FIG. 3. Surface reconstruction (a) before; (b) after. Blue balls represent atoms on the topmost layer, light blue the second layer, red the third layer, and gray the fourth layer.

We note that surface reconstruction has taken place on Zn $\{10\bar{1}2\}$ surface, and the change of the first interlayer spacing appears very large. The agreement between quantum mechanics calculations and R-EAM based calculations in Figure 2 hints that the agreement is also valid in the surface reconstruction. Indeed, both calculations give nearly identical reconstructions of Zn $\{10\bar{1}2\}$, as shown in Figure 3; the EAM based calculations lead to no reconstruction. Here we emphasize that the R-EAM based calculations give the same surface reconstruction as quantum mechanics calculations, and deemphasize the comparison between R-EAM based and EAM based calculations. The de-emphasis is due to the fact that this particular EAM potential should not well describe configurations of low coordination due to the use of electron density as a non-monotonic function of distance, as shown in [10]. In the reconstruction process, not only the first two layers of atoms in the topmost double-layer align into one layer, the third layer (which is from the second double-layer) also aligns into the same surface layer. As a result of this reconstruction the rumpled Zn $\{10\bar{1}2\}$ surface becomes much smoother, with atomic packing as a mixture of $\{0001\}$ and $\{10\bar{1}1\}$ surfaces; the $\{0001\}$ surface has hexagonal symmetry of atomic packing, and the $\{10\bar{1}1\}$ surface has alternating triangular and square symmetry of atomic packing.

Before closing, we note that other approaches have been proposed to improve the EAM. By introducing angular dependence [26], the modified EAM aims to describe covalent solids. Through the use of the electron density matrix instead of a scalar [27], another approach aims to describe angular dependent bonding of alloys. Through inclusion of both electron density and its gradient, the third approach resembles the improvement from local density approximation [25] to generalized gradient approbation [20] in density functional theory calculations. The fourth approach includes Columbic interactions so as to describe ionic solids [28]. None of these approaches take into account the response in the R-EAM.

In summary, we have proposed the concept of R-EAM and demonstrated its advantages over EAM. Conceptually, this method effectively takes into account the difference of electrons that each atom contributes as the atom's environment changes. The computational cost using R-EAM potentials is comparable to that using EAM potential, within a factor of two. In contrast to the modest increase of computational cost, the advantages of R-EAM over EAM are significant when it comes to the description of elastic constants and surface properties. The R-EAM does not impose the three constraints on elastic constants that the EAM does. The R-EAM based calculation results agree with quantum mechanics results on surfaces – in terms of interlayer spacing and surface reconstruction; EAM based calculations do not agree. The second advantage of the R-EAM – that is, its description of surfaces – makes it particularly relevant to predictive simulations of nanostructured materials.

**Acknowledgement:** The authors acknowledge financial support from the Department of Energy Office of Basic Energy Sciences (DE-FG02-09ER46562) for surface studies, and the Defense Threat Reduction Agency (HDTRA1-09-1-0027) for studies of HCP metals. They also thank Dr. Yuri Mishin for providing confirmation on the EAM value of surface formation energy of Ti {0001}.